\documentstyle[preprint,aps]{revtex}
\begin{document}
\preprint{IFT-P.037/93; DOE/ER/40427-12-N93}
\def\overlay#1#2{\setbox0=\hbox{#1}\setbox1=\hbox to \wd0{\hss
#2\hss}#1\hskip -2\wd0\copy1}
\def\lsim{\mathrel{\rlap{\lower4pt\hbox{\hskip1pt$\sim$}}
    \raise1pt\hbox{$<$}}}         %less than or approx. symbol
\def\gsim{\mathrel{\rlap{\lower4pt\hbox{\hskip1pt$\sim$}}
    \raise1pt\hbox{$>$}}}         %greater than or approx. symbol
\title{\vspace{5.0cm}{Self-consistent solution of the Schwinger-Dyson
equations}\\
{for the nucleon and meson propagators}}
\author{{M.E. Bracco{\thanks{E-mail: mirian@ift.uesp.ansp.br}},
 A. Eiras{\thanks{E-mail: aeiras@ift.uesp.ansp.br}},
 G. Krein{\thanks{E-mail: gkrein@ift.uesp.ansp.br}}}\\
{Instituto de F\'{\i}sica Te\'orica - Universidade Estadual Paulista}\\
{Rua Pamplona, 145 - 01405-900 S\~ao Paulo - Brazil}\\
and\\
{L. Wilets{\thanks{E-mail: wilets@alpher.npl.washington.edu}}}\\
{Department of Physics FM-15, University of Washington,
Seattle WA 98195 - USA}}
\maketitle
\newpage
\begin{abstract}
The Schwinger-Dyson equations for the nucleon and meson propagators are
solved self-consistently in an approximation that goes beyond the
Hartree-Fock approximation. The traditional approach consists in solving the
nucleon Schwinger-Dyson equation with bare meson propagators and bare
meson-nucleon vertices; the corrections to the meson propagators are
calculated using the bare nucleon propagator and bare nucleon-meson vertices.
It is known that such an approximation scheme produces the appearance of
ghost poles in the propagators. In this paper the coupled system of
Schwinger-Dyson equations for the nucleon and the meson propagators are
solved self-consistently including vertex corrections. The interplay of
self-consistency and vertex corrections on the ghosts problem is investigated.
It is found that the self-consistency does not affect significantly the
spectral properties of the propagators. In particular, it does not affect the
appearance of the ghost poles in the propagators.
\end{abstract}
\newpage
\section{Introduction}
The development of relativistic many-body theories for the nucleus is one of
the most important goals of contemporary nuclear theory. Models based on the
methods of relativistic quantum field theory have been developed for more than
two decades.

The starting point for understanding the many-nucleon problem is a
description of the elementary processes in vacuum:  the nucleon propagator,
meson-nucleon scattering, and the $N-N$ interaction.  Successes and
difficulties with relativistic meson-nucleon field theory have been the
subject of papers for more than half a century.  We will certainly not
detail the history here, but note that a nagging inconsistency in
(almost) all calculations has been the appearance of ghost poles.

Brown, Puff and Wilets \cite{bpw}, for example, calculated the nucleon
propagator by summing all planar meson diagrams with one nucleon line
using $\pi$-, $\rho$-, and $\omega$-mesons.  No cutoffs were introduced.
The renormalized nucleon propagator was well-defined and self-consistent,
but contained a pair of conjugate complex poles located approximately one
GeV off the real and imaginary axes.  The full propagator, including these
unphysical poles, was used with some success to describe the isovector nucleon
magnetic moment,  pi-nucleon scattering\cite{NuttWilets}, and nucleon-nucleon
scattering\cite{Nutt}.  (The last did require cut-offs in the N-N interaction,
but yielded better chi-squared fits to scattering data with fewer parameters
than the then current Paris potential).  The inclusion of the complex poles
was essential. Nevertheless, the occurence of the complex poles remained an
enigma.

Several interpretations of the appearance of the poles have been proffered,
including the statement that it is a  signal of the inconsistency of any
local, relativistic field theory, and that a field theory with asymptotic
freedom (e.g. QCD) is required.

The program of the previous section was driven by the
interpretation that the appearance of the ghosts is an artifact of the
approximations, and that progressively better calculations should lead to
the receding or elimination of the ghosts, but that for consistency one
must keep the ghosts as they emerge from the calculations at each stage.

Another interpretation is that it is an effective theory, and that one
should be prepared to introduce further parameters to ensure physical
quantities.

In a recent paper\cite{knpw}, the problem of ghosts poles in the nucleon
propagator was investigated. The appearance of the ghost
poles is related to the short distance behavior of the model
interactions\cite{bpw}; asymptotically free theories appear to be free of
ghost poles\cite{pe}. An interesting possibility to eliminate the complex
poles is the regularization of the theory by means of vector meson dressing
of nucleon-meson vertices. It is known that in a theory with neutral vector
mesons there are vertex corrections that generate a strongly damped vertex
function in the ultraviolet\cite{known}. In quantum electrodynamics, such
corrections give rise to the Sudakov form factor\cite{sud}. When the
Sudakov form factor, generated by massive vector mesons, is included in the
HF approximation to the Schwinger-Dyson equation (SDE) for the nucleon
propagator, the ghost poles disappear. A similar result was obtained by
Allendes and Serot\cite{allse} earlier in the study of the ghost pole in the
meson propagator. Those authors concluded that the Sudakov corrected
propagator is free of ghost poles.

It is the purpose of the present paper to solve self-consistently
the coupled system of Schwinger-Dyson equations (SDE) for the nucleon and
meson propagators and investigate the role of self-consistentency on the
appearance of ghost poles in the propagators. Vertex corrections are introduced
by means of form factors.

There is an extensive literature on calculations of nuclear matter and finite
nuclei properties based on the Walecka scalar-vector model\cite{wa}. In
general, the applications have been performed using Hartree-Fock(HF) type
of approximations. In a relativistic HF approximation, the single-nucleon
propagator is calculated by solving self-consistently the (SDE) using bare
meson propagators and bare meson-nucleon vertices. An additional approximation
has been the neglect of the quantum vacuum of the nucleon propagator. The
corrections to the meson propagators are usually calculated considering the
vacuum polarization correction using nucleon propagators with an effective
mass. Although the nucleon propagator is solved self-consistently by means of
the SDE, the self-consistency is only partial, since the meson propagators
used are the bare ones. The meson propagators satisfy their own SDE, which
require for their solution the nucleon propagator. A self-consistent solution
requires the consideration of the coupled system of nucleon and meson SDE's.

Besides the lack of self-consistency, the neglect of the quantum vacuum in
the nucleon sector is a major limitation. It is exactly the nontrivial
nature of the vacuum of a relativistic quantum field theory that motivates
the introduction of models which go beyond the usual nonrelativistic
approach. However, severe difficulties arise in including the vacuum
effects beyond the one-loop Hartree approximation. The inclusion of these
vacuum corrections leads to catastrophic results due to the presence of the
ghost poles in the propagators. Among other things, the ghosts lead to a
large imaginary part to the nuclear matter energy.

The paper is organized as follows. In section II we present the model for
the interacting nucleon-meson system. We briefly review the spectral
representation of the propagators and their inverses and discuss the
renormalization procedure. In section III we discuss the coupled system of
Schwinger-Dyson equations for the nucleon and meson propagators in terms of
their spectral representations. Section IV presents the method of solution
of the equations and presents our numerical results. Conclusions are
presented in section V.

\section{The model}
In this paper we consider a model field theory with nucleons
($\psi$), pions ($\vec \pi$), and vector isoscalar mesons ($\omega$). The
model Lagrangian density is

\begin{eqnarray}
{\cal L} = &&{\bar \psi}(i \gamma_{\mu} \partial^{\mu}
-i g_{0\pi}\gamma_5 \vec \tau \cdot \vec \pi -
g_{0\omega} \gamma_{\mu} \omega^{\mu}) \psi \nonumber \\
&&- {1\over 4} F_{\mu \nu} F^{\mu \nu}
- {1\over 2} m_{\omega}^2 \omega_{\mu} \omega^{\mu}
+{1\over 2} \partial_{\mu} \vec \pi \cdot \partial^{\mu} \vec \pi
-{1\over 2} m_{\pi}^2 \vec \pi \cdot \vec \pi\;,
\label{lag}
\end{eqnarray}

\noindent
where
$F^{\mu \nu} = \partial^{\mu} \omega^{\nu} - \partial^{\nu} \omega^{\mu}$.

As usual, the nucleon propagator is defined by

\begin{equation}
G_{\alpha \beta}(x'-x)=-i<0|T[\psi_{\alpha}(x')\bar \psi_{\beta}(x)]|0>\;,
\label{defnucpro}
\end{equation}

\noindent
where $|0>$ represents the physical vacuum state. The $\pi$- and $\omega$-meson
propagators are defined respectively by

\begin{equation}
D_{\pi}^{ij}(x'-x)=-i<0|T[\pi^i(x') \pi^j(x)]|0>\;,
\label{defpipro}
\end{equation}

\noindent
and

\begin{equation}
D_{\omega}^{\mu \nu}(x'-x)=-i<0|T[\omega^{\mu}(x') \omega^{\nu}(x)]|0>\;.
\label{defompro}
\end{equation}

The Schwinger-Dyson equations for the nucleon and meson propagators in momentum
space are given by the following expressions, Fig. 1,

\noindent
(a) nucleon:

\begin{equation}
G(p)=G^{(0)}(p)+G^{(0)}(p)\Sigma(p)G(p)\;,
\label{sdenuc}
\end{equation}

\begin{eqnarray}
\Sigma(p)&=&-3 i g_{0\pi}^2 \int {d^4q\over (2\pi)^4} \gamma_5
D_{\pi}(q^2)G(p-q)\Gamma_5(p-q,p;q) \nonumber \\
&+&i g_{0\omega}^2 \int {d^4q\over (2\pi)^4} \gamma_{\mu}
D_{\omega}^{\mu \nu}(q^2)G(p-q)\Gamma_{\nu}(p-q,p;q)\;,
\label{nuceq}
\end{eqnarray}

\noindent
(b) pion:

\begin{equation}
D_{\pi}^{ij}(q^2)={D_{\pi}^{(0)}}^{ij}(q^2)+
{D_{\pi}^{(0)}}^{ik}(q^2)\Pi_{\pi}^{kl}(q^2)D_{\pi}^{lj}(q^2)\;,
\label{sdepi}
\end{equation}

\begin{equation}
\Pi_{\pi}^{ij}(q^2)= ig_{0\pi}^2 \int {d^4p \over (2\pi)^4}
Tr[\gamma_5 \tau^i G(p)\Gamma_5^j(p, p+q; q) G(p+q)]\;.
\label{pieq}
\end{equation}

\noindent
(c) omega:

\begin{equation}
D_{\omega}^{\mu \nu}(q^2)={D_{\omega}^{\mu \nu}}^{(0)}(q^2)
+{D_{\omega}^{\mu \rho}}^{(0)}(q^2)\Pi_{\omega}^{\rho \sigma}(q^2)
D_{\omega}^{\sigma \nu}(q^2)\;,
\label{sdeom}
\end{equation}

\begin{equation}
\Pi_{\omega}^{\mu \nu}(q^2)=- ig_{0\omega}^2 \int {d^4p \over (2\pi)^4}
Tr[\gamma^{\mu} G(p) \Gamma^{\nu}(p, p+q; q) G(p+q)]\;.
\label{omeq}
\end{equation}

In the above equations, $\Gamma_5^i(p, p+q; q)$ and $\Gamma^{\mu}(p, p+q; q)$
are the three-point $\pi$-nucleon and $\omega$-nucleon vertex functions,
respectively. They satisfy their own Schwinger-Dyson equations. These relate
the three-point functions to four-point vertices and so on {\it ad infinitun}.
In practice one has to truncate this infinite set. In this paper we truncate
the SDE's by postulating a specific form for the three-point functions (see
below).

Next, we discuss the spectral representations of the propagators and their
inverses. We do not intend to review the subject of spectral representations,
we simply make use of the relevant equations for the purposes of the present
paper. We refer the reader to {~Refs.~\cite{Roman}~{-}~\cite{wil}} for
an extensive discussion on the subject. Let us start with the nucleon
propagator. The spectral representation of the nucleon propagator (in momentum
space) can be written as

\begin{equation}
G(p)=\int_{- \infty}^{+ \infty} d\kappa\; {A(\kappa) \over {{\not\!p} - \kappa
+ i\epsilon}}\;.
\label{kaleh}
\end{equation}

\noindent
$A(\kappa)$ is the spectral function.  It represents the probability that a
state of mass $|\kappa|$ is created by $\psi$ or $\bar \psi$, and as such it
must be non-negative. Negative $\kappa$ corresponds to states with
opposite parity to the nucleon.

Defining the projection operators

\begin{equation}
P_{\pm}(p)={1\over 2}\left(1 \pm {{\not\!p} \over w_p}\right)\;,
\end{equation}

\noindent
where

\begin{eqnarray}
w_p = \sqrt{p^2} =
          \left\{ \begin{array}{ll}
                     \sqrt{p^2}, & \mbox{if $p^2 > 0$} \\
                    i\sqrt{-p^2}, & \mbox{if $p^2 < 0$},
                              \end{array}
                              \right.
\label{wp}
\end{eqnarray}

\noindent
$G(p)$ can be rewritten conveniently as

\begin{equation}
G(p)=P_{+}(p)\tilde G (w_p+i\epsilon)+P_{-}(p)\tilde G(-w_p-i\epsilon)\;,
\label{gpro}
\end{equation}

\noindent
where $\tilde G(z)$, $z=\pm(w_p+i\epsilon)$, is given by the dispersion
integral

\begin{equation}
\tilde G(z) = \int_{- \infty}^{+ \infty} d\kappa\;
{A(\kappa) \over {z - \kappa}}\;.
\label{gtil}
\end{equation}

\noindent
The inverse of the propagator can also be written in terms of the projection
operators $P_{\pm}(p)$ as

\begin{equation}
G^{-1}(p)=P_{+}(p)\tilde G^{-1} (w_p+i\epsilon)+
P_{-}(p)\tilde G^{-1}(-w_p-i\epsilon)\;.
\label{ginvpro}
\end{equation}

Since $A(\kappa)$ is supposed to be non-negative, it is simple to show that
$\tilde G(z)$ can have no poles or zeros off the real axis. This is
known as the Herglotz property. Now, if $\tilde G(z)$ possesses the Herglotz
property, then so does $\tilde G^{-1}(z)$. This permits us to write a
spectral representation for $\tilde G^{-1}(z)$,

\begin{eqnarray}
\tilde G^{-1}(z)&=&z-M_0 - \tilde \Sigma(z)\nonumber \\
&=&z-M_0 - \int_{- \infty}^{+ \infty} d\kappa
{T(\kappa) \over {z-\kappa}}\;.
\label{nucself}
\end{eqnarray}

\noindent
The function $\tilde \Sigma(z)$ is related to the $\Sigma(q)$ of
Eq. (\ref{sdenuc}) by the projection operators $P_{\pm}(q)$ as in
Eq. (\ref{ginvpro}). If $T(\kappa)$ is non-negative, the Herglotz
property is also satisfied for $\tilde G^{-1}(z)$.

In general, the integral in Eq. (\ref{nucself}) needs renormalization. The
usual
mass and wave-function renormalizations are performed by imposing the condition
that the renormalized propagator has a pole at the physical nucleon mass $M$,
with unit residue. This implies that the renormalized propagator
$\tilde G_{R}(z)$, defined as

\begin{equation}
\tilde G_{R}(z) \equiv \tilde G(z)/ Z_2\;,
\label{defgr}
\end{equation}

\noindent
is given by the following expression:

\begin{equation}
\tilde G_R(z) = \int_{- \infty}^{+\infty} d\kappa\; {A_{R}(\kappa) \over
{z-\kappa}}\;.
\label{rg}
\end{equation}

\noindent
The renormalized inverse is given by

\begin{equation}
\tilde G_{R}^{-1}(z) =(z-M)\left[ 1-(z-M)\int_{- \infty}^{+\infty} d\kappa\;
{T_{R}(\kappa) \over {(\kappa-M)^2(z-\kappa)}}\right]\;.
\label{rself}
\end{equation}

\noindent
In the above expressions, $A_{R}(\kappa) = A(\kappa)/Z_2$ and
$T_{R}(\kappa) = Z_2T(\kappa)$. In terms of renormalized quantities, $Z_2$ can
be written as

\begin{eqnarray}
Z_2 &=& 1 - \int_{- \infty}^{+\infty} d\kappa {T_{R}(\kappa)
\over{(\kappa-M)^2}}\label{z2t}\\
&=&
\left[\int_{- \infty}^{+\infty} d\kappa A_{R}(\kappa)\right]^{-1}\;.
\label{z2a}
\end{eqnarray}

The spectral functions $A_R(\kappa)$ and $T_R(\kappa)$ are related by

\begin{eqnarray}
A_R(\kappa)&=&\delta(\kappa-M)+
|{\tilde G}_R^{-1}(\kappa(1+i\epsilon))|^{-2}T_R(\kappa)\label{AandT} \\
&\equiv &\delta(\kappa-M)+\bar A_R(\kappa)\;.
\label{Abar}
\end{eqnarray}

Let us now consider the spectral representations of the meson propagators.
The isospin structure of the $\pi$-meson propagator is such that
$D_{\pi}^{ij}(q^2)=\delta^{ij}D_{\pi}(q^2)$. For $D_{\pi}(q^2)$ one can write
the spectral representation

\begin{equation}
D_{\pi}(z)=\int_0^{\infty} d\sigma^2 \;{\rho_{\pi}(\sigma^2) \over
{z - \sigma^2 }}\;,
\label{disppi}
\end{equation}

\noindent
where $\rho_{\pi}(\sigma^2)$ is the pion spectral function. It represents the
probability that a state of mass $\sqrt{\sigma^2}$ is created by the pion
field and as such it must be non-negative. The meaning of the
complex variable $z$ is that the physical propagator $D_{\pi}(q^2)$ is the
limit of $D_{\pi}(z)$ when $z \rightarrow q^2 + i\epsilon$.

Using the SDE for the pion, Eq. (\ref{sdepi}), the inverse of $D_{\pi}(z)$ can
be written in terms of the pion self-energy $\Pi(z)$ as

\begin{equation}
D_{\pi}^{-1}(z)=z-{m_{\pi}^0}^2-\Pi_{\pi}(z)\;.
\label{invpiprop}
\end{equation}

\noindent
Similarly to the case of the nucleon, one can write a spectral representation
for $D_{\pi}^{-1}(z)$

\begin{equation}
D_{\pi}^{-1}(z)=z-{m_{\pi}^0}^2-\int_{0}^{\infty} d\sigma^2\;
{S_{\pi}(\sigma^2) \over {z-\sigma^2} }\;.
\label{Cauchypi}
\end{equation}

The renormalized propagator is again obtained by fixing the pole position
at the physical mass, and the residue at the pole equal to one. The
renormalized propagator $D_{\pi R}(z)$, defined as

\begin{equation}
D_{\pi R}(z) \equiv D_{\pi}(z)/ Z_{3 \pi}\;,
\label{rgm}
\end{equation}

\noindent
is then

\begin{equation}
D_{\pi R}(z)=\int_0^{\infty} d\sigma^2 \;{\rho_{\pi R}(\sigma^2) \over
{z - \sigma^2 }}\;.
\label{Rdisppi}
\end{equation}

\noindent
Its inverse is given by

\begin{equation}
D_{\pi R}^{-1}(z)=(z-m_{\pi}^2)\left[1-(z-m_{\pi}^2)
\int_{0}^{\infty} d\sigma^2\;
{ S_{\pi R}(\sigma^2) \over
{(\sigma^2-m_{\pi}^2)^2(z-\sigma^2)} }\right]\;.
\label{RCauchypi}
\end{equation}

\noindent
The renormalized spectral functions are defined as $\rho_{\pi R}(\sigma^2) =
\rho_{\pi}(\sigma^2)/Z_{3 \pi}$ and $S_{\pi R}(\sigma^2) =
Z_{3 \pi}S_{\pi}(\sigma^2)$.

In terms of the renormalized quantities, $Z_{3 \pi}$ is given by

\begin{eqnarray}
Z_{3 \pi}&=& 1-\int_{0}^{\infty} d\sigma^2\;
{ S_{\pi R}(\sigma^2) \over
{(\sigma^2-m_{\pi}^2)^2} } \label{Z3Pi}  \\
&=&\left[\int_{0}^{\infty}d\sigma^2\;
\rho_{\pi R}(\sigma^2)\right]^{-1}
\label{Z3rho}\;.
\end{eqnarray}

The spectral functions $\rho_{\pi R}$ and $S_{\pi R}$ are related by

\begin{eqnarray}
\rho_{\pi R}(q^2)&=&\delta (q^2-m_{\pi}^2)+|D_{\pi R}^{-1}|^2
S_{\pi R}(q^2)\label{RhoandPi} \\
&\equiv &\delta (q^2-m_{\pi}^2) + \bar \rho_{\pi R}(q^2)\;.
\label{rhobar}
\end{eqnarray}

Let us now consider the $\omega$-meson propagator. Since the baryon current is
conserved the $\omega$-meson self-energy $\Pi_{\omega}^{\mu \nu}(q^2)$, must
satisfy

\begin{equation}
q_{\mu} \Pi_{\omega}^{\mu \nu}(q^2)=
q_{\nu} \Pi_{\omega}^{\mu \nu}(q^2)=0\;.
\label{ccons}
\end{equation}

\noindent
Therefore, the Lorentz structure of $\Pi_{\omega}^{\mu \nu}$ must be

\begin{equation}
\Pi_{\omega}^{\mu \nu}(q^2)=\left(g^{\mu \nu}-q^{\mu} q^{\nu}/q^2\right)
\Pi_{\omega}(q^2)\;.
\label{Lstruct}
\end{equation}

Substituting this in the SDE for $\omega$-meson propagator, Eq. (\ref{sdeom}),
$D_{\omega}^{\mu \nu}(q^2)$ can be written as

\begin{equation}
D_{\omega}^{\mu \nu}(q^2)=-g^{\mu \nu}D_{\omega}(q^2)\;,
\label{defomeg}
\end{equation}

\noindent
where

\begin{equation}
D_{\omega}(q^2)={1 \over
{q^2-{m_{\omega}^0}^2-\Pi_{\omega}(q^2)+i \epsilon} }\;.
\label{omprop}
\end{equation}

\noindent
Terms proportional to $q^{\mu} q^{\nu}$ in Eq. (\ref{defomeg}) can be neglected
when using $D_{\omega}^{\mu \nu}$ in Eq. (\ref{nuceq}), because of current
conservation.

The spectral representation of $D_{\omega}$ is

\begin{equation}
D_{\omega}(z)=\int_0^{\infty} d\sigma^2 \;{\rho_{\omega}(\sigma^2) \over
 {z - \sigma^2}}\;.
\label{dispom}
\end{equation}

\noindent
As in the case of the pion, one can write the Cauchy representation for the
inverse of the $\omega$-meson propagator as

\begin{equation}
D_{\omega}^{-1}(z)=z-{m_{\omega}^0}^2-\int_{0}^{\infty}
d\sigma^2\; {S_{\omega}(\sigma^2) \over {z-\sigma^2} }\;.
\label{Cauchyom}
\end{equation}

Renormalization proceeds as for the pion. The renormalized quantities
are given by expressions similar to the ones for the pion,
Eqs. (\ref{Rdisppi}, \ref{RCauchypi}, \ref{Z3rho}, \ref{rhobar}), with the
$\pi$ indices replaced by $\omega$ indices.

\section{Schwinger-Dyson equations}

We start with the nucleon SDE, Eq. (\ref{sdenuc}). It can be written as

\begin{equation}
G^{-1}(p)=G^{-1}(p)-\Sigma(p)\;,
\label{sdeinvnuc}
\end{equation}

\noindent
where $\Sigma(p)$ is given by Eq. (\ref{nuceq}). To proceed, we need to specify
the form of the vertex functions  $\Gamma_5^i(p, p+q; q)$ and
$\Gamma^{\mu}(p, p+q; q)$. In the usual HF approximation,
$\Gamma_5^i(p, p+q; q)=\tau^i \gamma_5$, and
$\Gamma^{\mu}(p, p+q; q)=\gamma^{\mu}$. In this paper we consider vertex
functions written as

\begin{eqnarray}
\Gamma_5^i(p_1, p_2; q)&=&\tau^i \gamma_5 F_5(p_1, p_2; q) \label{Gam5} \\
\Gamma^{\mu}(p_1, p_2; q)&=&\gamma^{\mu} F_V(p_1, p_2; q) \label{GamV}\;,
\end{eqnarray}

\noindent
where $F_5(p_1, p_2; q)$ and $F_V(p_1, p_2; q)$ are scalar functions.

Substituting Eqs. (\ref{Gam5},\ref{GamV}) and the spectral representations for
$G(q)$, $D_{\pi}$ and $D_{\omega}$ in the integral for $\Sigma(q)$,
Eq. (\ref{nuceq}), and applying the projection operators $P_{\pm}(p)$ to
Eq. (\ref{sdeinvnuc}), one obtains:

\begin{equation}
T_R(\kappa)=\int_{-\infty}^{+\infty} d\kappa'K(\kappa,\kappa')A_R(\kappa')
\label{tkap}
\end{equation}

\noindent
where $K(\kappa,\kappa')$ is given by

\begin{eqnarray}
K(\kappa, \kappa') &=& K_{\pi}(\kappa, \kappa'; m_{\pi}^2) +
2 K_{\omega}(\kappa, \kappa'; m_{\omega}^2) \nonumber \\
&+& \int_0^{\infty} d\sigma^2 \bar \rho_{\pi R}(\sigma^2)
K_{\pi}(\kappa, \kappa'; \sigma^2)+
2 \int_0^{\infty} d\sigma^2 \bar \rho_{\omega R}(\sigma^2)
K_{\omega}(\kappa, \kappa'; \sigma^2)\;.
\label{K}
\end{eqnarray}

\noindent
$K_{\pi}(\kappa, \kappa'; m^2)$ and $K_{\omega}(\kappa, \kappa'; m^2)$ are
respectively the $\pi$-nucleon and $\omega$-nucleon scattering kernels

\begin{eqnarray}
K_{\pi}(\kappa, \kappa'; m^2)&=& 3\left({g_{\pi}\over 4\pi}\right)^2
\left[\kappa^4-2\kappa^2({\kappa'}^2+m^2)+({\kappa'}^2
-m^2)^2\right]^{1/2}\nonumber \\
&\times&{1\over 2|\kappa|^3}\left[(\kappa-\kappa')^2-m^2\right]
\theta(\kappa^2-(|\kappa'|+m)^2)F_5(\kappa, \kappa'; m)\;,
\label{Kpi}
\end{eqnarray}

\noindent
and

\begin{eqnarray}
K_{\omega}(\kappa, \kappa'; m^2)&=& \left({g_{\omega}\over 4\pi}\right)^2
\left[\kappa^4-2\kappa^2({\kappa'}^2+m^2)+({\kappa'}^2
-m^2)^2\right]^{1/2}\nonumber \\
&\times&{1\over 2|\kappa|^3}\left[(\kappa-\kappa')^2-2\kappa \kappa'-m^2\right]
\theta(\kappa^2-(|\kappa'|+m)^2)F_V(\kappa, \kappa'; m)\;.
\label{Komega}
\end{eqnarray}

\noindent
$\bar \rho_{\pi R}(\sigma^2)$ is related to $S_{\pi R}$ as shown in
Eqs. (\ref{RhoandPi}-\ref{rhobar}) (similarly for
$\bar \rho_{\omega R}(\sigma^2)$).

The meson self-energies,  $S_{\pi R}(q^2)$ and $S_{\omega R}(q^2)$, are
obtained
using the spectral representation of the nucleon propagator in
Eqs. (\ref{pieq}, \ref{omeq}). $S_{\pi R}(q^2)$ is given by:

\begin{eqnarray}
S_{\pi R}(q^2)&=&S_{\pi}(M,M;q^2)  \nonumber \\
&+& 2 \int_{-\infty}^{\infty} d\kappa \bar{A}_R(\kappa)
S_{\pi}(M,\kappa;q^2) \nonumber\\
&+& \int_{-\infty}^{\infty} d\kappa d\kappa' \bar{A}_R(\kappa)
\bar{A}_R(\kappa')S_{\pi}(\kappa,\kappa';q^2)\;,
\label{Pipi}
\end{eqnarray}

\noindent
where $\bar A(\kappa)$ is defined in Eq. (\ref{Abar}), and

\begin{eqnarray}
S_{\pi}(\kappa,\kappa';q^2)&=&\left({g_{\pi}^2 \over 4 \pi^2}
\right)
\biggl[1-{(\kappa-\kappa')^2 \over q^2}\biggr]
[q^4 - 2 q^2(\kappa ^2+\kappa'^2)+(\kappa^2-\kappa'^2)^2]^{1 \over 2}
\nonumber \\
& & \times \Theta (q^2 -(|\kappa|+|\kappa'|)^2)F_5(\kappa,\kappa';q)\;,
\label{imPipi}
\end{eqnarray}

\noindent
with $g_{\pi}^2=Z_2 g_{0\pi}$ and $g_{\omega}^2=Z_2 g_{0\omega}$.

For the $\omega$-meson, we have the same expression as in Eq. (\ref{Pipi}),
with the indice $\pi$ replaced by $\omega$ and

\begin{eqnarray}
S_{\omega}(\kappa,\kappa';q^2) &=&\left({g_{\omega}^2 \over 8 \pi^2}
\right) \biggl\{1-{(\kappa-\kappa')^2 \over q^2}+
{1\over 3q^4}[q^4 - 2 q^2(\kappa ^2+\kappa'^2)+(\kappa^2-\kappa'^2)^2]
\biggr\} \nonumber \\
& &\times[q^4 - 2 q^2(\kappa ^2+\kappa'^2)+(\kappa^2-\kappa'^2)^2]^{1 \over 2}
\Theta (q^2 -(|\kappa|+|\kappa'|)^2)F_V(\kappa,\kappa';q)\;.
\label{imPiom}
\end{eqnarray}

\section{Numerical results}

The problem now consists in  finding $A_R(\kappa)$, $\rho_{\pi R}(\sigma^2)$,
and $\rho_{\omega R}(\sigma^2)$.
The strategy adopted is to solve the  equations by iteration; it proceeds as
follows:

\noindent
1) start solving for $A_R$ with the bare $\pi$ and $\omega$ propagators
($\bar \rho_{\pi R}=\bar \rho_{\omega R}=0$). This is the usual Hartree-Fock
solution for the nucleon propagator including vertex corrections by means of
the form factors of Eqs. (\ref{Gam5}, \ref{GamV})\cite{knpw};

\noindent
2) with this $A_R$, obtain $S_{R \pi}(q^2)$ and
$S_{R \omega}(q^2)$ with Eq. (\ref{Pipi});

\noindent
3) then use Eq. (\ref{rhobar}) to obtain $\bar \rho_{\pi R}$ and similarly
$\bar \rho_{\omega R}$;

\noindent
4) use these $\bar \rho_{\pi R}$ and $\bar \rho_{\omega R}$ to obtain
new $K(\kappa, \kappa')$;

\noindent
5) with this $K(\kappa, \kappa')$, solve for $A_R$, go back to 2),
and cycle to convergence.

Initially, we considered bare vertices: $F_5(p_1, p_2, q)=F_V(p_1, p_2, q)=1$,
and investigated the role of the self-consistency on the spectral functions.
We used the following values for the coupling constants and masses:

\begin{eqnarray}
{g_{\pi}^2 \over 4\pi}&=&14.6 \hspace{3.0cm} m_{\pi}=0.144\;M \\
\label{piconst}
{g_{\omega}^2 \over 4\pi}&=&6.36 \hspace{3.0cm} m_{\omega}=0.833\;M\;,
\label{omconst}
\end{eqnarray}

\noindent
where $M$ is the nucleon mass.

The converged spectral functions $A_R$, $\rho_{\pi R}$, and
$\rho_{\omega R}$ are shown (without the delta functions) in Figs. 2-4.
The solid (dashed) lines represent the self-consistent (not self-consistent)
solutions. The self-consistency does not affect the fermion spectral function
perceptively, but it affects the meson functions, although not very
importantly.
It is interesting to note that the effect of the self-consistency is opposite
in
$\rho_{\pi R}$ and $\rho_{\omega R}$; it increases the former and decreases the
last.

The role of the self-consistency on the appearance of ghost poles is shown
in Table 1. The not self-consistent meson spectral functions are the ones
obtained by calculating the nucleon loop in Figs. 1(b),1(c)
using the bare nucleon propagators ($\bar A_R=0$), i.e., these are the first
order  perturbative spectral functions. Clearly, the self-consistency does not
change much the position of the poles and residues of the nucleon propagator,
although it changes somewhat the ones of the meson propagators.

As discussed in Refs. \cite{bpw}, \cite{knpw}, the signal for the presence of
ghosts in the nucleon propagator is revealed by the fact that the
renormalization constant $Z_2$  calculated via the spectral function
of the nucleon self-energy, $T_R(\kappa)$, gives $Z_2=-\infty$.  From this
result one should have $Z_2^{-1}=0$. However, in order to get the integral of
$A_R$ equal to zero (see Eq. (\ref{z2a})) one has then to include the pair of
complex conjugated poles in $\tilde G_R$ (the real parts of the residues are
negative (see Table 1)). In the case of the renormalization constants of the
$\pi$ and $\omega$ mesons, we obtain exactly the same result: the $Z_3$'s
calculated via the spectral function of the self-energy, $S_R$,
gives $Z_3 \longrightarrow -\infty$, whereas to obtain zero for the integral
over the spectral function of the propagator, $\rho_R$, the residue of the
ghost pole has to be included.

In Ref. \cite{knpw}, the problem of ghosts poles in the nucleon
propagator was investigated using form factors at the nucleon-meson
vertices. Two types of form factors were used: (a) a Sudakov form factor,
which is generated by vector meson dressing of the vertices and (b) a
phenomenological form factor, of the monopole type. The conclusion there was
that both types of form factors are able to kill the ghosts.  However, as
remarked in that reference, a proper extension of the Sudakov form factor
to lower momenta is necessary for a better study of these issues. In this paper
we use only the simple monopole form factor to investigate the
interplay of self-consistency and vertex corrections on the spectral functions.
As in Ref. \cite{knpw}, we use for $F_5(p_1,p_2,q)$ and $F_V(p_1,p_2,q)$ the
following expressions:

\begin{equation}
F_5(p_1,p_2,q)=F_V(p_1,p_2,q)=
{1\over{1+|p_1^2/\Lambda^2|}}\;{1\over{1+|q^2/\Lambda^2|}}\;
{1\over{1+|p_2^2/\Lambda^2|}}\;,
\label{ff}
\end{equation}

\noindent
where $\Lambda$ is an ultraviolet cuttoff.

The modifications on the spectral functions due to the form factors can be
seen in Figs. 5-7, where we plotted $A_R$, $\rho_{\pi R}$, and
$\rho_{\omega R}$  (again without the delta functions) for two typical values
of cutoffs, $\Lambda=M$ (solid), $1.25\;M$ (long-dashed). The (short-dashed)
curves correspond to the case of $\Lambda=\infty$. The effect of the form
factor is to increase $A_R(\kappa)$ for negative $\kappa$, a result already
found in \cite{knpw}, and to increase (decrease) $\rho_{R \pi}$
($\rho_{R \omega}$). In Ref. \cite{knpw}, it was found that for a
$\Lambda < \Lambda_{crit} \approx 1.75\;M$ the ghost poles in the nucleon
propagator disappear. In the present case, we found that the self-consistency
does not alter significantly this value; for $\Lambda \lsim 1.60\;M$, the
ghosts of all propagators disappear.

We have also investigated the effect of the self-consistency on the
ghosts-free spectral functions, i.e. for several $\Lambda$'s smaller than
$\Lambda_{crit}$, we compared the self-consistent and not self-consistent
spectral functions. Surprisingly, the effect of the self-consistency is
neglegible, and is almost invisible when one plots the spectral functions.

Although on physical grounds one expects that the cutoffs for the $\pi$ and
$\omega$ vertices have different values, we used the same value
for both, since in this work we are mostly interested in the qualitative
effects. The consequences of the modifications induced by the form factors
on physical observables deserves a separate study. Work in this direction is
in progress.

\section{Conclusions and perspectives}

In this paper we have solved self-consistently the coupled set of
Schwinger-Dyson equations for the nucleon and $\pi$ and $\omega$ mesons in the
vacuum. The set of equations was truncated by postulating a three-point
meson-nucleon vertex function. The understanding of the vacuun properties of
the nucleon and meson propagators is a necessary first step towards the study
of the properties of nucleon and meson in nuclear matter, as well
as those of nuclear matter and finite nuclei. Although many of such properties
have been studied using relativistic quantum field models, the vacuun
polarization effects in medium have invariably been neglected.

The main conclusion of our investigation is the surprising result that the
self-consistency does not modify significantly the spectral properties of
the propagators. The appearance or disappearance of the ghost poles in
the propagators is not affected by the self-consistency.

One important aspect regarding the vacuum of meson-nucleon effective theories
that was not yet satisfactorily investigated is the role of the three-point
meson-nucleon vertex functions. In particular, the interplay of the infrared
and ultraviolet sectors of the $\omega$-nucleon three-point vertice is
extremely
important to the problem of ghosts poles, as shown in the recent studies of
Refs. (\cite{allse}, \cite{knpw}). Work in this direction is in progress.

The effects of the self-consistency on the nucleon and meson propagators in
nuclear matter, in connections to the problem of ghost poles remains an
open problem, although work in this direction has recently been
communicated\cite{korpa}.

\section{Acknowledgements}
The work of MEB, AE and GK was partially supported by the Brazilian
agencies CNPq and FAPESP. The work of LW was partially supported by the
U.S. Department of Energy.
\figure{FIG. 1. Diagrammatic representation of the Schwinger-Dyson equations
for the full (a)nucleon, (b)pion, and (c)omega propagators. The solid, wavy and
dashed lines represent respectively the nucleon, the $\omega$, and the $\pi$.
The blobs represent full propagators and vertices.}
\figure{\noindent FIG. 2. Self-consistent(solid) and not self-consistent
(dashed
   ) nucleon spectral function $A_R(\kappa)$. $\kappa$ is in units of the
nucleo
   n mass
$M$ and $A_R(\kappa)$ is in units of $M^{-1}$. The curves are multiplied by 5
for negative $\kappa$.}
\figure{\noindent FIG. 3. Self-consistent(solid) and not self-consistent
(dashed) $\pi$ spectral function $\rho_{\pi R}(\sigma^2)$. $\sigma^2$ is in
units of $M^2$ and $\rho_{\pi R}(\sigma^2)$ is in units of $M^{-2}$.
\figure{\noindent FIG. 4. Self-consistent(solid) and not self-consistent
(dashed) $\omega$ spectral function $\rho_{\omega R}(\sigma^2)$. The units
are the same as in Fig. 3.}
\figure{\noindent FIG. 5. $A_R(\kappa)$ for different values of the cutoff:
$\Lambda=M$ (solid), $1.25\;M$ (long-dashed), $\infty$ (short-dashed). Units
are the same as in Fig. 2. The short-dashed curve is multiplied by 5 for
negative $\kappa$.}
\figure{\noindent FIG. 6. $\rho_{\pi R}(\sigma^2)$ for same $\Lambda$'s as
in Fig. 5. Units are the same as in Fig. 3.}
\figure{\noindent FIG. 7. Same as in Fig. 6 for
$\rho_{\omega R}(\sigma^2)$.}
\begin{table}
\caption{Ghost poles. The first value is the pole position
and the second is the residue at the pole. The nucleon poles are in units
of $M$ and of the mesons are in units of $M^2$. }
\vspace{2.0mm}
\begin{tabular}{cccc}
${}    $ &  Self-consistent &  Not self-consistent   \\\hline
$N     $ &  $1.06\pm 1.25i\hspace{1.0cm}-0.77 \pm 0.20i $ &
$1.05 \pm 1.26i\hspace{1.0cm}-0.77 \pm 0.20i $\\
$\pi   $ &  $-1.04\hspace{1.0cm}-1.08$ &
$-1.44\hspace{1.0cm}-1.13                      $\\
$\omega$ &  $-3.50\hspace{1.0cm}-1.30$ &
$-5.68 \hspace{1.0cm}-1.49                     $\\
\end{tabular}
\end{table}
\end{document}